# Closed Form Approximation Solutions for the Restricted Circular Three Body Problem

Abu Bakr Mehmood, S. Umer Abbas and Ghulam Shabbir

Faculty of Engineering Sciences,

GIK Institute of Engineering Sciences and Technology

Topi, Swabi, NWFP, Pakistan

Email: shabbir@giki.edu.pk

**Abstract:**

An approach is developed to find approximate solutions to the restricted circular three body problem. The solution is useful in approximately describing the position vectors of three spherically symmetric masses, one of which has a much smaller mass than the other two. These masses perform free motion under each others' gravitational influence. The set of solutions is found using the lambert's wave function.

**Introduction**

The aim of this paper is to find approximations for the restricted circular three body problem. The problem by definition is to describe the free motions of three masses, two of which have spherically symmetric mass distributions and one of which is small compared to the other two. We develop a new approach in this paper, other approaches can be found in [1-6]. The smaller mass should be small enough in comparison to the other two, so that it can be approximated as a point mass. A typical real life application of the problem would be the motion of a probe between the earth and moon. Moreover, the spherically symmetric mass distributions of the earth and moon would allow them to be approximated by point masses. In the problem considered, it is further assumed that the motion of the two larger masses, say $m_1$ and $m_2$ is not affected by the presence, or motion of $m_3$, the smaller mass. It therefore follows that $m_1$ and $m_2$ execute two body motion under each other's gravitational influence only, whereas $m_3$ executes motion which is effected by both the presence and motion if $m_1$ and $m_2$. The motion of $m_1$ and $m_2$ shall be solved for, only by considering two body motion, and the motion of $m_3$, shall then be solved for by the use of generalized three body motion equations that we will soon develop. Figure 1 presents a diagrammatic representation of our system of three bodies, which form an isolated system in free space.

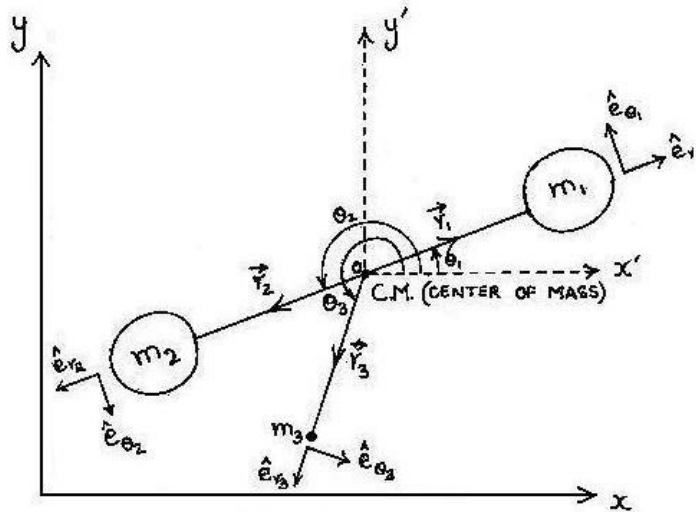

**Figure 1**

Our aim is to find $\mathbf{r}_1$, $\mathbf{r}_2$ and $\mathbf{r}_3$ explicitly as time functions. We choose to solve the problem in two dimensions. The bodies perform translational motion under each other's gravitational attraction. Since our assumptions allow us to approximate the three bodies as point masses, we neglect rotational motion.

**MAIN RESULTS**

It can be shown that the center of mass of the system moves with constant velocity, which can be found. We therefore attach an inertial frame of reference to the center of mass point, $ox'y'$, and model the system by the use of Newton's Law of Gravitational attraction. Before we proceed to model the system, a few points are worth discussion, presented as follows.

The mass $m_3$ is much smaller than $m_1$ and $m_2$ so that the presence or motion of $m_3$ has no influence on the motion of $m_1$ and $m_2$. Therefore the motion of $m_1$ and

$m_2$ is effected only by each other's presence and motion, and that of $m_3$ is effected by the presence and motion of both $m_1$ and $m_2$. It therefore follows that the Center of Mass of the system should be approximately unaffected, whether or not $m_3$ is present. $m_1$ and $m_2$ therefore execute two body motion, the motion of $m_3$ being effected by the motion and presence of $m_1$ and $m_2$. Also it follows that the Center of Mass should be approximately present on a straight line joining $m_1$ and $m_2$.

It should follow from the above arguments that

$$\theta_2 = \theta_1 + \pi \quad \forall t \quad \Rightarrow \dot{\theta}_2 = \dot{\theta}_1 \quad \text{and also}$$

$$\hat{e}_{r_1} = -\hat{e}_{r_2} \quad \text{and} \quad \hat{e}_{\theta_1} = -\hat{e}_{\theta_2} \quad \forall t$$

Where $\hat{e}_{r_1}$ and $\hat{e}_{r_2}$ are unit vectors in the direction of $\mathbf{r}_1$ and $\mathbf{r}_2$, respectively and where $\hat{e}_{\theta_1}$ and $\hat{e}_{\theta_2}$ are unit vectors perpendicular to $\hat{e}_{r_1}$ and $\hat{e}_{r_2}$, respectively. It should be noted that "." denotes $d/dt$. We now go on to model the system, by the use of the Newton's Law of Gravitational attraction. For the moment we relax the assumptions that $m_3 \ll m_1$ and $m_3 \ll m_2$ so that we first model the generalized three body problem. This model would later be utilized when we try describing the motion of $m_3$. Figure 2 presents the configuration of the bodies with the assumptions relaxed.

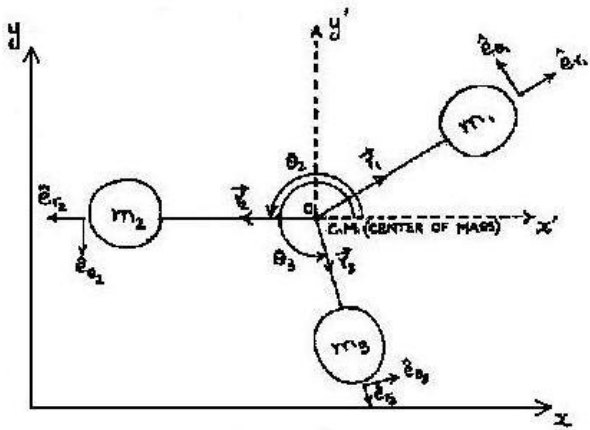

**Figure 2**

Modeling the system by considering the fact that the net force on each body is due to

the gravitational attraction of the other two bodies we get the following equations

$$\ddot{\mathbf{r}}_1 = \left(\frac{Gm_2}{|\mathbf{r}_2 - \mathbf{r}_1|^2}\right)\left[\frac{\hat{e}_{r_2} - \hat{e}_{r_1}}{|\hat{e}_{r_2} - \hat{e}_{r_1}|}\right] + \left(\frac{Gm_3}{|\mathbf{r}_3 - \mathbf{r}_1|^2}\right)\left[\frac{\hat{e}_{r_3} - \hat{e}_{r_1}}{|\hat{e}_{r_3} - \hat{e}_{r_1}|}\right] \quad (1a)$$

$$\ddot{\mathbf{r}}_2 = -\left(\frac{Gm_1}{|\mathbf{r}_2 - \mathbf{r}_1|^2}\right)\left[\frac{\hat{e}_{r_2} - \hat{e}_{r_1}}{|\hat{e}_{r_2} - \hat{e}_{r_1}|}\right] + \left(\frac{Gm_3}{|\mathbf{r}_3 - \mathbf{r}_2|^2}\right)\left[\frac{\hat{e}_{r_3} - \hat{e}_{r_2}}{|\hat{e}_{r_3} - \hat{e}_{r_2}|}\right] \quad (1b)$$

$$\ddot{\mathbf{r}}_3 = -\left(\frac{Gm_1}{|\mathbf{r}_3 - \mathbf{r}_1|^2}\right)\left[\frac{\hat{e}_{r_3} - \hat{e}_{r_1}}{|\hat{e}_{r_3} - \hat{e}_{r_1}|}\right] - \left(\frac{Gm_2}{|\mathbf{r}_3 - \mathbf{r}_2|^2}\right)\left[\frac{\hat{e}_{r_3} - \hat{e}_{r_2}}{|\hat{e}_{r_3} - \hat{e}_{r_2}|}\right] \quad (1c)$$

where $\hat{e}_{r_3}$ is a unit vector in the direction of $\mathbf{r}_3$. We now try removing the absolute value expressions

$$|\mathbf{r}_2 - \mathbf{r}_1|^2 = r_1^2 + r_2^2 - 2r_1 r_2 \cos(\theta_2 - \theta_1) \quad (2a)$$

Similarly,

$$|\mathbf{r}_3 - \mathbf{r}_1|^2 = r_1^2 + r_3^2 - 2r_1 r_3 \cos(\theta_3 - \theta_1) \quad (2b)$$

$$|\mathbf{r}_3 - \mathbf{r}_2|^2 = r_2^2 + r_3^2 - 2r_2 r_3 \cos(\theta_3 - \theta_2) \quad (2c)$$

Also, we can write

$$|\hat{e}_{r_2} - \hat{e}_{r_1}| = \sqrt{2}\sqrt{1 - \cos(\theta_2 - \theta_1)} \quad (3a)$$

Similarly

$$|\hat{e}_{r_3} - \hat{e}_{r_1}| = \sqrt{2}\sqrt{1 - \cos(\theta_3 - \theta_1)} \quad (3b)$$

and

$$|\hat{e}_{r_3} - \hat{e}_{r_2}| = \sqrt{2}\sqrt{1 - \cos(\theta_3 - \theta_2)} \quad (3c)$$

Using the above results in equations (1a), (1b) and (1c), we get

$$\ddot{\mathbf{r}}_1 = \left(\frac{Gm_2}{\sqrt{2}[r_1^2 + r_2^2 - 2r_1 r_2 \cos(\theta_2 - \theta_1)]\sqrt{1 - \cos(\theta_2 - \theta_1)}}\right)[\hat{e}_{r_2} - \hat{e}_{r_1}] +$$

$$\left(\frac{Gm_3}{\sqrt{2}[r_1^2 + r_3^2 - 2r_1 r_3 \cos(\theta_3 - \theta_1)]\sqrt{1 - \cos(\theta_3 - \theta_1)}}\right)[\hat{e}_{r_3} - \hat{e}_{r_1}] \quad (4a)$$

$$\ddot{\mathbf{r}}_2 = -\left(\frac{Gm_1}{\sqrt{2}[r_1^2 + r_2^2 - 2r_1 r_2 \cos(\theta_2 - \theta_1)]\sqrt{1 - \cos(\theta_2 - \theta_1)}}\right)[\hat{e}_{r_2} - \hat{e}_{r_1}]$$

$$+ \left(\frac{Gm_3}{\sqrt{2}[r_2^2 + r_3^2 - 2r_2 r_3 \cos(\theta_3 - \theta_2)]\sqrt{1 - \cos(\theta_3 - \theta_2)}}\right)[\hat{e}_{r_3} - \hat{e}_{r_2}] \quad (4b)$$

$$\ddot{\mathbf{r}}_3 = -\left(\frac{Gm_1}{\sqrt{2}[r_1^2 + r_3^2 - 2r_1r_3\cos(\theta_3 - \theta_1)]\sqrt{1-\cos(\theta_3 - \theta_1)}}\right)[\hat{e}_{r_3} - \hat{e}_{r_1}]$$

$$-\left(\frac{Gm_2}{\sqrt{2}[r_2^2 + r_3^2 - 2r_2r_3\cos(\theta_3 - \theta_2)]\sqrt{1-\cos(\theta_3 - \theta_2)}}\right)[\hat{e}_{r_3} - \hat{e}_{r_2}] \quad (4c)$$

Now we can resolve each equation in polar coordinates to derive the required scalar differential equations. Before we do that, we must resolve the unit vectors in terms of each other. Figure 3 serves as an aid while performing this task.

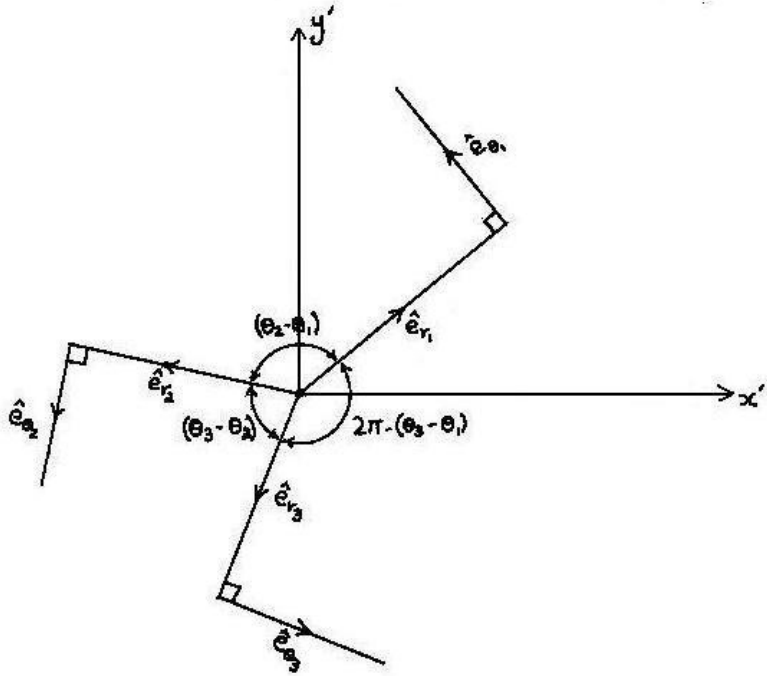

**Figure 3**

Resolving $\hat{e}_{r_1}$, $\hat{e}_{r_2}$ and $\hat{e}_{r_3}$ we get

$$\hat{e}_{r_2} = \cos(\theta_2 - \theta_1)\hat{e}_{r_1} + \sin(\theta_2 - \theta_1)\hat{e}_{\theta_1}$$

$$\hat{e}_{r_3} = \cos(\theta_3 - \theta_1)\hat{e}_{r_1} + \sin(\theta_3 - \theta_1)\hat{e}_{\theta_1}$$

$$\hat{e}_{r_1} = \cos(\theta_2 - \theta_1)\hat{e}_{r_2} + \sin(\theta_2 - \theta_1)\hat{e}_{\theta_2}$$

$$\hat{e}_{r_3} = \cos(\theta_3 - \theta_2)\hat{e}_{r_2} + \sin(\theta_3 - \theta_2)\hat{e}_{\theta_2}$$

$$\hat{e}_{r_1} = \cos(\theta_3 - \theta_1)\hat{e}_{r_3} + \sin(\theta_3 - \theta_1)\hat{e}_{\theta_3}$$

$$\hat{e}_{r_2} = \cos(\theta_3 - \theta_2)\hat{e}_{r_3} + \sin(\theta_3 - \theta_2)\hat{e}_{\theta_3}$$

Resolving equations (4a), (4b) and (4c) parallel and perpendicular to $\hat{e}_{r_1}$, $\hat{e}_{r_2}$ and $\hat{e}_{r_3}$, respectively and comparing coefficients of the unit vectors on both sides for each equation, we can derive the following necessary scalar differential equations

$$\ddot{r}_1 - r_1 \dot{\theta}_1^2 = \frac{Gm_2[\cos(\theta_2 - \theta_1) - 1]}{\sqrt{2}[r_1^2 + r_2^2 - 2r_1 r_2 \cos(\theta_2 - \theta_1)]\sqrt{1 - \cos(\theta_2 - \theta_1)}} +$$

$$\frac{Gm_3[\cos(\theta_3 - \theta_1) - 1]}{\sqrt{2}[r_1^2 + r_3^2 - 2r_1 r_3 \cos(\theta_3 - \theta_1)]\sqrt{1 - \cos(\theta_3 - \theta_1)}} \quad (5a)$$

$$r_1 \ddot{\theta}_1 + 2\dot{r}_1 \dot{\theta}_1 = \frac{Gm_2[\sin(\theta_2 - \theta_1)]}{\sqrt{2}[r_1^2 + r_2^2 - 2r_1 r_2 \cos(\theta_2 - \theta_1)]\sqrt{1 - \cos(\theta_2 - \theta_1)}} +$$

$$\frac{Gm_3[\sin(\theta_3 - \theta_1)]}{\sqrt{2}[r_1^2 + r_3^2 - 2r_1 r_3 \cos(\theta_3 - \theta_1)]\sqrt{1 - \cos(\theta_3 - \theta_1)}} \quad (5b)$$

$$\ddot{r}_2 - r_2 \dot{\theta}_2^2 = \frac{Gm_1[\cos(\theta_2 - \theta_1) - 1]}{\sqrt{2}[r_1^2 + r_2^2 - 2r_1 r_2 \cos(\theta_2 - \theta_1)]\sqrt{1 - \cos(\theta_2 - \theta_1)}} +$$

$$\frac{Gm_3[\cos(\theta_3 - \theta_2) - 1]}{\sqrt{2}[r_2^2 + r_3^2 - 2r_2 r_3 \cos(\theta_3 - \theta_2)]\sqrt{1 - \cos(\theta_3 - \theta_2)}} \quad (6a)$$

$$r_2 \ddot{\theta}_2 + 2\dot{r}_2 \dot{\theta}_2 = -\frac{Gm_1[\sin(\theta_2 - \theta_1)]}{\sqrt{2}[r_1^2 + r_2^2 - 2r_1 r_2 \cos(\theta_2 - \theta_1)]\sqrt{1 - \cos(\theta_2 - \theta_1)}}$$

$$-\frac{Gm_3[\sin(\theta_3 - \theta_2)]}{\sqrt{2}[r_2^2 + r_3^2 - 2r_2 r_3 \cos(\theta_3 - \theta_2)]\sqrt{1 - \cos(\theta_3 - \theta_2)}} \quad (6b)$$

$$\ddot{r}_3 - r_3 \dot{\theta}_3^2 = \frac{Gm_1[\cos(\theta_3 - \theta_1) - 1]}{\sqrt{2}[r_1^2 + r_3^2 - 2r_1 r_3 \cos(\theta_3 - \theta_1)]\sqrt{1 - \cos(\theta_3 - \theta_1)}} +$$

$$\frac{Gm_2[\cos(\theta_3 - \theta_2) - 1]}{\sqrt{2}[r_2^2 + r_3^2 - 2r_2 r_3 \cos(\theta_3 - \theta_2)]\sqrt{1 - \cos(\theta_3 - \theta_2)}} \quad (7a)$$

$$r_3 \ddot{\theta}_3 + 2\dot{r}_3 \dot{\theta}_3 = -\frac{Gm_1[\sin(\theta_3 - \theta_1)]}{\sqrt{2}[r_1^2 + r_3^2 - 2r_1 r_3 \cos(\theta_3 - \theta_1)]\sqrt{1 - \cos(\theta_3 - \theta_1)}}$$

$$-\frac{Gm_2[\sin(\theta_3 - \theta_2)]}{\sqrt{2}[r_2^2 + r_3^2 - 2r_2 r_3 \cos(\theta_3 - \theta_2)]\sqrt{1 - \cos(\theta_3 - \theta_2)}} \quad (7b)$$

One may ask as to why we derived all these equations in this form. For convenience of the reader we restate here that this model of the generalized three body problem will be useful when it comes to describing the free motion of $m_3$ under the influence of $m_1$ and $m_2$.

In the restricted case then, **r**₁ and **r**₂ can be found explicitly as time functions by considering only two body motion (the motion of $m_1$ and $m_2$ under each others' influence). Finding **r**₁ and **r**₂ as time functions would mean that we find $r_1(t)$, $\theta_1(t)$, $r_2(t)$ and $\theta_2(t)$. Having done this, we can find $r_3(t)$ and $\theta_3(t)$ by elimination from equations $(5a)$, $(5b)$, $(6a)$ and $(6b)$. Here is the summary of our methodology for the restricted case

**1:** Find $r_1(t)$, $\theta_1(t)$, $r_2(t)$ and $\theta_2(t)$ by considering two body motion only (that of $m_1$ and $m_2$).

**2:** Find $r_3(t)$ and $\theta_3(t)$ by elimination from equations $(5a)$ through $(6b)$.

We do not hesitate in pointing out that another option could be the simultaneous solution of equations $(7a)$ and $(7b)$ for $r_3(t)$ and $\theta_3(t)$, but as should be obvious, this approach would be quite cumbersome and perhaps not possible.

We now go on to find **r**₁ and **r**₂ by considering two body motion only. The configuration is shown in figure 4.

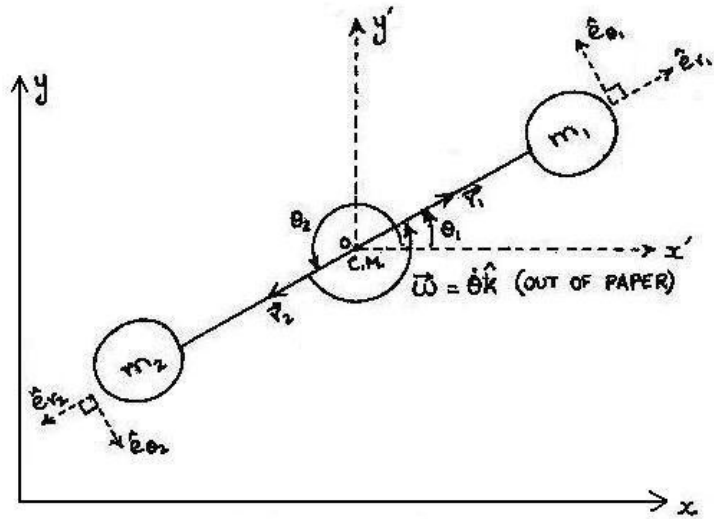

**Figure 4**

Here $ox'y'$ is an inertial frame of reference attached to the center of mass. Doubtlessly $ox'y'$ is inertial reference frame, since we can prove that the Center of Mass of the system has a constant velocity for all $t$. Modeling the system by using Newton's

Gravitational Law, we can derive the following equations

$$\ddot{\mathbf{r}}_1 = -\left[\frac{Gm_2}{(r_1+r_2)^2}\right]\hat{e}_{r_1} \tag{8a}$$

and

$$\ddot{\mathbf{r}}_2 = -\left[\frac{Gm_1}{(r_1+r_2)^2}\right]\hat{e}_{r_2} \tag{8b}$$

The above equations are derived by making use of $ox'y'$ as reference frame. For the two body motion and the restricted three body case, the following equations are valid

$$\hat{e}_{r_1} = -\hat{e}_{r_2} \tag{9a}$$

$$\hat{e}_{\theta_1} = -\hat{e}_{\theta_2} \tag{9b}$$

Subtracting (8a) from (8b) while making use of (9a) we get

$$\ddot{\mathbf{r}}_2 - \ddot{\mathbf{r}}_1 = -\left[\frac{G(m_1+m_2)}{(r_1+r_2)^2}\right]\hat{e}_{r_2} \tag{10}$$

We now define $\mathbf{r} = \mathbf{r}_2 - \mathbf{r}_1$, $\mathbf{r} = r\hat{e}_r$, $r = r_1 + r_2$, $\hat{e}_r = \hat{e}_{r_2}$ and $\hat{e}_\theta = \hat{e}_{\theta_2}$. Where $\hat{e}_r$ is a unit vector along $\mathbf{r}$ and $\hat{e}_\theta$ is a unit vector perpendicular to $\hat{e}_r$. Incorporating the above definitions in (10) we get

$$\ddot{\mathbf{r}} = -\left[\frac{G(m_1+m_2)}{r^2}\right]\hat{e}_r \tag{11}$$

Resolving $\ddot{\mathbf{r}}$ in polar coordinates we have

$$\left(\ddot{r} - r\dot{\theta}^2\right)\hat{e}_r + \left(r\ddot{\theta} + 2\dot{r}\dot{\theta}\right)\hat{e}_\theta = -\left[\frac{G(m_1+m_2)}{r^2}\right]\hat{e}_r$$

Here $\theta(t)$ is the rotation angle of $\mathbf{r}$. It also follows that $\theta(t) = \theta_2(t)$. Comparing coefficients of $\hat{e}_r$ and $\hat{e}_\theta$ in the above relation, we find

$$\ddot{r} - r\dot{\theta}^2 = -\left[\frac{G(m_1+m_2)}{r^2}\right] \tag{12a}$$

$$r\ddot{\theta} + 2\dot{r}\dot{\theta} = 0 \tag{12b}$$

Multiplying (12b) by $r$ and integrating on both sides we get

$$r^2\dot{\theta} = h = constant \tag{13}$$

Now for solving equation (12a), we make use of the substitution $r = \dfrac{1}{u}$ and relation (13) to get

$$r(\theta) = \frac{1}{\left[C\cos(\theta+\phi) + \left(\frac{G(m_1+m_2)}{h^2}\right)\right]} \quad (14)$$

where $C$ and $\phi$ are constants of integration, determinable by the incorporation of initial conditions. Substituting (14) in (13), using separation of variables and integrating on both sides, we can show that

$$\frac{2}{\left[\left(\frac{G(m_1+m_2)}{h^2}\right) - C\right]^2} \left[\left(\frac{k-1}{4k^{\frac{3}{2}}}\right) \ln\left[\frac{(\tan 0.5\theta - \sqrt{k})(\tan 0.5\theta_o + \sqrt{k})}{(\tan 0.5\theta + \sqrt{k})(\tan 0.5\theta_o - \sqrt{k})}\right]\right] -$$

$$\left(\frac{k+1}{4k}\right)\left[\frac{1}{\tan 0.5\theta + \sqrt{k}} + \frac{1}{\tan 0.5\theta - \sqrt{k}} - \frac{1}{\tan 0.5\theta_o + \sqrt{k}} - \frac{1}{\tan 0.5\theta_o - \sqrt{k}}\right]$$

$$= h(t - t_o)$$

It is apparent from this expression that we cannot find $\theta(t)$ explicitly. Had this been possible, we could try finding $r(t)$ by substituting $\dot{\theta}(t)$ into equation (13). Having found $r(t)$ and $\theta(t)$, our next step would be to attempt to find $\mathbf{r}_1(t)$ and $\mathbf{r}_2(t)$ from equations (8a) and (8b) by making use of the relation $\theta(t) = \theta_2(t) = \theta_1(t) + \pi$. However to our disappointment, this is rendered impossible by the inability to find $\theta(t)$ explicitly from the above expression.

We now suggest a simple mathematical argument, for angular velocities that are small ($|\dot{\theta}| < 1$ *radians per second*) and for $r$ not very large ($r \ll \pm\infty$) we claim that $\dot{\theta}^2 \simeq 0 \Rightarrow r\dot{\theta}^2 \simeq 0$. It follows that $\ddot{r} - r\dot{\theta}^2 \simeq \ddot{r}$ and equation (12a) reduces to

$$\ddot{r} \simeq -\left(\frac{G(m_1+m_2)}{r^2}\right) \forall \, |\dot{\theta}| < 1 \, rad/s \, \& \, |r| \ll \infty \quad (15)$$

Multiplying (12a) by $\dot{r} \, dt$ and integrating on the left hand side with respect to $t$ and on the right hand side with respect to $r$, we get

$$\frac{dr}{dt} = \pm\left[\frac{A}{r} + B\right]^{\frac{1}{2}}$$

where $A = 2G(m_1+m_2)$ and $B = \dot{r}_o^2 - \frac{2G(m_1+m_2)}{r_o}$. Separating variables and

integrating gives

$$\int_{r_0}^{r} \left[\frac{A}{Br}+1\right]^{-\frac{1}{2}} dr = \pm\sqrt{B} \int_{t_o}^{t} dt \tag{16}$$

where $r_o$ is the initial value of $r$ at initial time $t_o$. If we try evaluating an exact value of the integral on the left hand side of the preceding equation, we can obtain an equation relating $r$ and $t$, but $r(t)$ cannot be explicitly solved for. We therefore try and keep things simple, by the use of a binomial approximation

$$\left[\left(\frac{A}{Br}\right)+1\right]^{-\frac{1}{2}} \simeq 1-\left(\frac{A}{2Br}\right) \quad \forall \, |r| > \left| \frac{2G(m_1+m_2)}{\dot{r}_o^2 - \left(\frac{2G(m_1+m_2)}{r_o}\right)} \right|$$

It should be noted that here that $B = 0$ is not an allowable value, however this will soon be taken care of. Replacing $\left[\left(\frac{A}{Br}\right)+1\right]^{-\frac{1}{2}}$ by its approximation in (16) and integrating on both sides we get

$$r + \ln r^{-k} = f(t) \tag{17}$$

where $k = \left(\frac{A}{2B}\right)$ and $f(t) = \pm\sqrt{B}(t-t_o) + r_o - \ln r_o^k$. Here we restrict $B$ to be necessarily positive, so that $f(t)$ is real. This of course, is one of our added solvability condition. Later we will collectively state all solvability conditions. Solving for $r(t)$ then, explicitly from (17) we get

$$r(t) = -k * lambertw\left[\frac{-e^{-\frac{f}{k}}}{k}\right] \tag{18}$$

where '$lambertw$' is the notation used for the lambert's wave function. Substituting the values of $k$ and $f(t)$ in (18), and performing few manipulations, we can derive the following expression

$$r(t) = -\left(\frac{A}{2B}\right) lambertw\left[c_4 e^{c_5 t}\right] \tag{19}$$

where $c_4 = c_1 e^{c_2 t_o}$, $c_5 = -c_2$, $c_1 = -\left(\frac{2B}{A}\right) e^{(-\frac{2B}{A})(r_o - \ln r_o^{\frac{A}{2B}})}$ and $c_2 = \pm\left(\frac{2B\sqrt{B}}{A}\right)$. Of course this approximation is valid only when the following conditions hold

$$\left| \frac{2G(m_1+m_2)}{r_o^2 - \frac{2G(m_1+m_2)}{r_o}} \right| < |r| << \infty \text{ and } |\dot{\theta}| < 1 \text{ rad}/s \ (\dot{\theta}_1 = \dot{\theta}_2 = \dot{\theta} \ \forall \ t)$$

Having derived $r(t)$, we now go on to derive $r_1(t)$ and $r_2(t)$. Consider again equations (8a) and (8b). We can derive (20) and (21) from (8a) and (8b) respectively, with a procedure similar to the one adopted in the derivation of (15) from (11)

$$\ddot{r}_1 \simeq -\frac{Gm_2}{r^2} \ \forall \ |\dot{\theta}_1| < 1 \text{ rad}/s \ \& \ r_1 << \pm\infty \tag{20}$$

$$\ddot{r}_2 \simeq -\frac{Gm_1}{r^2} \ \forall \ |\dot{\theta}_2| < 1 \text{ rad}/s \ \& \ r_2 << \pm\infty \tag{21}$$

Note that the conditions required for (20) and (21) to hold true, follow automatically from the conditions required for (15) to hold true. It should be noted that $r(t)$ is now known in both (20) and (21). All we need to do is merely substitute for $r(t)$ into (20) and (21) and integrate twice with respect to time to get the expressions

$$r_1(t) = k_a t - k_a t_o +$$
$$\left(\frac{k_b}{c_5}\right)\left[\frac{1}{2}(lambertw[c_4 e^{c_5 t}])^4 + (lambertw[c_4 e^{c_5 t}])^3 + \frac{1}{2}(lambertw[c_4 e^{c_5 t}])^2\right]$$
$$-\left(\frac{k_b}{c_5}\right)\left[\frac{1}{2}(lambertw[c_4 e^{c_5 t_o}])^4 + (lambertw[c_4 e^{c_5 t_o}])^3 + \frac{1}{2}(lambertw[c_4 e^{c_5 t_o}])^2\right] + r_{10}$$

(22)

where $r_{1o} = r_1(0)$, $\dot{r}_{1o} = \dot{r}_1(0)$, $k_a = \dot{r}_{1o} - \left(\frac{k_1}{2c_5}\right)\left[1 + 2lambertw(c_4 e^{c_5 t_o})\right]$, $k_b = \left(\frac{k_1}{2c_5}\right)$ and

$$k_1 = -\left(\frac{4B^2 Gm_2}{A^2}\right).$$

And

$$r_2(t) = k_c t - k_c t_o +$$
$$\left(\frac{k_d}{c_5}\right)\left[\frac{1}{2}(lambertw[c_4 e^{c_5 t}])^4 + (lambertw[c_4 e^{c_5 t}])^3 + \frac{1}{2}(lambertw[c_4 e^{c_5 t}])^2\right]$$
$$-\left(\frac{k_d}{c_5}\right)\left[\begin{array}{c}\frac{1}{2}(lambertw[c_4 e^{c_5 t_o}])^4 + \\ (lambertw[c_4 e^{c_5 t_o}])^3 + \frac{1}{2}(lambertw[c_4 e^{c_5 t_o}])^2\end{array}\right] + r_{20} \tag{23}$$

where $r_{2o} = r_2(0)$, $\dot{r}_{2o} = \dot{r}_2(0)$, $k_c = \dot{r}_{2o} - \left(\dfrac{k_2}{2c_2}\right)\left[1 + 2lambertw(c_4 e^{c_5 t_o})\right]$, $k_d = \left(\dfrac{k_2}{2c_2}\right)$

and $k_2 = -\left(\dfrac{4B^2 G m_1}{A^2}\right)$.

We will now present the formal procedure for deriving $\theta_1(t)$ and $\theta_2(t)$. Multiplying (12b) by $r$, we can derive the form $\frac{d}{dt}[r^2 \dot{\theta}] = 0 \Rightarrow d(r^2 \dot{\theta}) = 0$. Integrating and solving for $\dot{\theta}(t)$ would then yield

$$\dot{\theta}(t) = r_o^2 \, \dot{\theta}_o \, r^{-2}(t) \tag{24}$$

Since $\theta(t) = \theta_2(t)$ and $\theta_2(t) = \theta_1(t) + \pi \Rightarrow \dot{\theta}(t) = \dot{\theta}_2(t) = \dot{\theta}_1(t) \;\; \forall \; t$. Now using these relations and equation (24) we get

$$\dot{\theta}_1 = r_o^2 \, \dot{\theta}_o \, r^{-2}(t) \tag{25}$$

$$\dot{\theta}_2 = r_o^2 \, \dot{\theta}_o \, r^{-2}(t) \tag{26}$$

Now, substituting the expression for $r(t)$ from equation (19), separating variables and integrating within the necessary limits, we can derive the relations

$$\theta_1(t) = \theta_{1o} + \left(\dfrac{4 r_o^2 \, \dot{\theta}_o \, B^2}{c_5 A^2}\right)(1 + 2lambertw\left[c_4 e^{c_5 t_o}\right])(lambertw\left[c_4 e^{c_5 t_o}\right])^2$$

$$- \left(\dfrac{4 r_o^2 \, \dot{\theta}_o \, B^2}{c_5 A^2}\right)(1 + 2lambertw\left[c_4 e^{c_5 t}\right])(lambertw\left[c_4 e^{c_5 t}\right])^2 \tag{27}$$

$$\theta_2(t) = \theta_{2o} + \left(\dfrac{4 r_o^2 \, \dot{\theta}_o \, B^2}{c_5 A^2}\right)(1 + 2lambertw\left[c_4 e^{c_5 t_o}\right])(lambertw\left[c_4 e^{c_5 t_o}\right])^2$$

$$- \left(\dfrac{4 r_o^2 \, \dot{\theta}_o \, B^2}{c_5 A^2}\right)(1 + 2lambertw\left[c_4 e^{c_5 t}\right])(lambertw\left[c_4 e^{c_5 t}\right])^2 \tag{28}$$

where, $\theta_{1o}$ and $\theta_{2o}$ are the values of $\theta_1(t)$ and $\theta_2(t)$ at time $t_o$, respectively.

Our task of finding $r_1(t)$, $r_2(t)$, $\theta_1(t)$ and $\theta_2(t)$ is essentially complete, results given in (22), (23), (27) and (28). We can now move on to find $r_3(t)$ and $\theta_3(t)$ by algebraic manipulations on equations (5a) through (6b), instead of solving (7a) and (7b) simultaneously. It should be noted that (5a) through (7b) represent the generalized situation, where $m_3$ has an effect on the motion of $m_1$ and $m_2$. Since we want to solve for the restricted case, we enforce our assumptions for the restricted case onto equations (5a) through (7b). This can easily be accomplished by replacing $\theta_2 - \theta_1$ by $\pi$ in all the six equations (since $\theta_2 = \theta_1 + \pi$ for the restricted circular three body problem). Imposing these conditions we get simpler equations

$$\ddot{r}_1 - r_1 \dot{\theta}_1^2 = \left( \frac{Gm_3 [\cos(\theta_3 - \theta_1) - 1]}{\sqrt{2}[r_1^2 + r_3^2 - 2r_1 r_3 \cos(\theta_3 - \theta_1)]\sqrt{1 - \cos(\theta_3 - \theta_1)}} \right)$$

$$- \left( \frac{Gm_2}{[r_1^2 + r_2^2 + 2r_1 r_2]} \right) \quad (29a)$$

$$r_1 \ddot{\theta}_1 + 2\dot{r}_1 \dot{\theta}_1 = \left( \frac{Gm_3 \sin(\theta_3 - \theta_1)}{\sqrt{2}[r_1^2 + r_3^2 - 2r_1 r_3 \cos(\theta_3 - \theta_1)]\sqrt{1 - \cos(\theta_3 - \theta_1)}} \right) \quad (29b)$$

$$\ddot{r}_2 - r_2 \dot{\theta}_2^2 = \left( \frac{Gm_3 [\cos(\theta_3 - \theta_2) - 1]}{\sqrt{2}[r_2^2 + r_3^2 - 2r_2 r_3 \cos(\theta_3 - \theta_2)]\sqrt{1 - \cos(\theta_3 - \theta_2)}} \right)$$

$$- \left( \frac{Gm_1}{[r_1^2 + r_2^2 + 2r_1 r_2]} \right) \quad (30a)$$

$$r_2 \ddot{\theta}_2 + 2\dot{r}_2 \dot{\theta}_2 = -\left( \frac{Gm_3 \sin(\theta_3 - \theta_2)}{\sqrt{2}[r_2^2 + r_3^2 - 2r_2 r_3 \cos(\theta_3 - \theta_2)]\sqrt{1 - \cos(\theta_3 - \theta_2)}} \right)$$

$$(30b)$$

$$\ddot{r}_3 - r_3 \dot{\theta}_3^2 = \left( \frac{Gm_1 [\cos(\theta_3 - \theta_1) - 1]}{\sqrt{2}[r_1^2 + r_3^2 - 2r_1 r_3 \cos(\theta_3 - \theta_1)]\sqrt{1 - \cos(\theta_3 - \theta_1)}} \right) +$$

$$\left( \frac{Gm_2 [\cos(\theta_3 - \theta_2) - 1]}{\sqrt{2}[r_2^2 + r_3^2 - 2r_2 r_3 \cos(\theta_3 - \theta_2)]\sqrt{1 - \cos(\theta_3 - \theta_2)}} \right) \quad (31a)$$

$$r_3 \ddot{\theta}_3 + 2\dot{r}_3 \dot{\theta}_3 = -\left( \frac{Gm_1 \sin(\theta_3 - \theta_1)}{\sqrt{2}[r_1^2 + r_3^2 - 2r_1 r_3 \cos(\theta_3 - \theta_1)]\sqrt{1 - \cos(\theta_3 - \theta_1)}} \right)$$

$$- \left( \frac{Gm_2 \sin(\theta_3 - \theta_2)}{\sqrt{2}[r_2^2 + r_3^2 - 2r_2 r_3 \cos(\theta_3 - \theta_2)]\sqrt{1 - \cos(\theta_3 - \theta_2)}} \right) \quad (31b)$$

It should be noted that $(7a)$ and $(7b)$ remain unaffected inspite of the imposition of our assumptions. The best idea is definitely to obtain $r_3(t)$ and $\theta_3(t)$ by manipulations on $(29a)$ to $(30b)$, instead of the simultaneous solution of $(31a)$ and $(31b)$.

We identify a time function $f_1(t)$ from equation $(29a)$, which is of the form

$$f_1(t) = \left( \frac{\cos(\theta_3 - \theta_1) - 1}{\sqrt{2[r_1^2 + r_3^2 - 2r_1 r_3 \cos(\theta_3 - \theta_1)]}\sqrt{1 - \cos(\theta_3 - \theta_1)}} \right) \qquad (32a)$$

Here $f_1(t)$ has been considered as a time function since we expect each of the right hand side variables to be solvable explicitly as time functions. Of these right hand side variables we have already found $r_1(t)$ and $\theta_1(t)$. Since $r_3$ and $\theta_3$ should also be time functions, it follows that $f_1$ is essentially a function of time. Substituting equation $(32a)$ in equation $(29a)$ and solving for $f_1(t)$ we get

$$f_1(t) = \left( \frac{\ddot{r}_1 - r_1 \dot{\theta}_1^2}{Gm_3} \right) + \left( \frac{\left(\frac{m_2}{m_3}\right)}{r_1^2 + r_2^2 + 2r_1 r_2} \right)$$

(we may choose $r_1 \dot{\theta}_1^2 \simeq 0$ since $|\dot{\theta}_1| < 1 \ rad \ / \ s \ \& \ r_1 << \pm\infty$, however it is better not to do so, for obtaining better approximations). Using the same idea we can rewrite $(29b)$ to $(30b)$ in simpler forms, as will be clearly demonstrated in what follows.

We identify the following time function from equation $(29b)$

$$f_2(t) = \left( \frac{\sin(\theta_3 - \theta_1) - 1}{\sqrt{2[r_1^2 + r_3^2 - 2r_1 r_3 \cos(\theta_3 - \theta_1)]}\sqrt{1 - \cos(\theta_3 - \theta_1)}} \right) \qquad (32b)$$

Substituting equation $(32b)$ in $(29b)$ and solving for $f_2(t)$ yields

$$f_2(t) = \left( \frac{r_1 \ddot{\theta}_1 + 2\dot{r}_1 \dot{\theta}_1}{Gm_3} \right)$$

Similarly, $f_3(t)$ is identified and solved for (as follows) by use of equation $(30a)$

$$f_3(t) = \left( \frac{\cos(\theta_3 - \theta_2) - 1}{\sqrt{2[r_2^2 + r_3^2 - 2r_2 r_3 \cos(\theta_3 - \theta_2)]}\sqrt{1 - \cos(\theta_3 - \theta_2)}} \right) \qquad (33a)$$

$$f_3(t) = \left(\frac{\ddot{r_2} - r_2 \dot{\theta_2}^2}{Gm_3}\right) + \left(\frac{\left(\frac{m_1}{m_3}\right)}{r_1^2 + r_2^2 + 2r_1 r_2}\right)$$

(we may choose $r_2 \dot{\theta_2}^2 \simeq 0$ since $|\dot{\theta_2}| < 1\ rad\ /\ s\ \&\ r_2 << \pm\infty$, however it is better not to do so, for reasons of better approximations.)

Similarly, we can choose $f_4(t)$, from equation (30b) to be the following

$$f_4(t) = \left(\frac{\sin(\theta_3 - \theta_2) - 1}{\sqrt{2[r_2^2 + r_3^2 - 2r_2 r_3 \cos(\theta_3 - \theta_2)]}\sqrt{1 - \cos(\theta_3 - \theta_2)}}\right) \quad (33b)$$

Now substitution of (33b) into (30b) gives the following relation

$$f_4(t) = \left(\frac{r_2 \ddot{\theta_2} + 2\dot{r_2}\dot{\theta_2}}{-Gm_3}\right)$$

From (32a) we may write

$$\sqrt{2[r_1^2 + r_3^2 - 2r_1 r_3 \cos(\theta_3 - \theta_1)]}\sqrt{1 - \cos(\theta_3 - \theta_1)} = \left(\frac{\cos(\theta_3 - \theta_1) - 1}{f_1(t)}\right) \quad (34a)$$

Putting (34a) in (32b), we get

$$f_2(t) = \left(\frac{\sin(\theta_3 - \theta_1)}{\cos(\theta_3 - \theta_1) - 1}\right) f_1(t)$$

which can be rewritten as

$$[f_2(t)]\cos(\theta_3 - \theta_1) + [-f_1(t)]\sin(\theta_3 - \theta_1) = f_2(t) \quad (34b)$$

We can write the above equation in a more compact form, as follows

$$k(t)\cos(\theta_3 - \theta_1 - \psi(t)) = f_2(t) \quad (34c)$$

where $k(t) = \pm\sqrt{f_1^2(t) + f_2^2(t)}$ and $\psi(t) = \tan^{-1}\left(\frac{-f_1(t)}{f_2(t)}\right)$. Solving for $\theta_3(t)$ from (34c) and replacing the expressions for $k(t)$ and $\psi(t)$ we get

$$\theta_3(t) = \cos^{-1}\left[\frac{f_2(t)}{\pm\sqrt{f_1^2(t) + f_2^2(t)}}\right] + \theta_1(t) + \tan^{-1}\left[\frac{-f_1(t)}{f_2(t)}\right] \quad (35)$$

It follows that we can write equations (32a) through (32b) in the following forms

$$r_3^2 + (-2r_1)r_3\cos(\theta_3 - \theta_1) + r_1^2 = \left(\frac{\cos(\theta_3 - \theta_1) - 1}{\sqrt{2}f_1[1 + \cos(\theta_3 - \theta_1)]^{\frac{1}{2}}}\right) \quad (36a)$$

$$r_3^2 + (-2r_1)r_3\cos(\theta_3 - \theta_1) + r_1^2 = \left(\frac{\sin(\theta_3 - \theta_1)}{\sqrt{2}f_2[1 - \cos(\theta_3 - \theta_1)]^{\frac{1}{2}}}\right) \quad (36b)$$

$$r_3^2 + (-2r_2)r_3\cos(\theta_3 - \theta_2) + r_2^2 = \left(\frac{\cos(\theta_3 - \theta_2) - 1}{\sqrt{2}f_3[1 - \cos(\theta_3 - \theta_2)]^{\frac{1}{2}}}\right) \quad (37a)$$

$$r_3^2 + (-2r_2)r_3\cos(\theta_3 - \theta_2) + r_2^2 = \left(\frac{\sin(\theta_3 - \theta_2)}{\sqrt{2}f_4[1 - \cos(\theta_3 - \theta_2)]^{\frac{1}{2}}}\right) \quad (37b)$$

Subtracting (37b) from (36b) and solving for $r_3(t)$, we can derive the expression

$$r_3(t) = \left[\frac{\left[\left(\frac{\sin(\theta_3 - \theta_1)}{\sqrt{2}f_2[1 - \cos(\theta_3 - \theta_1)]^{\frac{1}{2}}}\right) - \left(\frac{\sin(\theta_3 - \theta_2)}{\sqrt{2}f_4[1 - \cos(\theta_3 - \theta_2)]^{\frac{1}{2}}}\right) + r_2^2 - r_1^2\right]}{2r_2\cos(\theta_3 - \theta_2) - 2r_1\cos(\theta_3 - \theta_1)}\right] \quad (38)$$

It should be noted that $r_3(t)$ has been found explicitly as a function of time, since all variables involved on the right hand side of (38) are known as time functions.

Our task is essentially complete since equations $(22), (23), (27), (28), (35)$ and (38) define explicit closed form approximations for $r_1(t), r_2(t), \theta_1(t), \theta_2(t), \theta_3(t)$ and $r_3(t)$ respectively, valid for the cases

$$|\mathbf{r}| = |\mathbf{r}_1| + |\mathbf{r}_2| > \left|\frac{2G(m_1 + m_2)}{\dot{r}_o^2 - \left(\frac{2G(m_1 + m_2)}{r_o}\right)}\right|$$

Or equivalently

$$|\mathbf{r}_1| + |\mathbf{r}_2| > \left|\frac{2G(m_1 + m_2)}{(\dot{r}_{1o} + \dot{r}_{2o})^2 - \left(\frac{2G(m_1 + m_2)}{r_{1o} + r_{2o}}\right)}\right|$$

Since

$$r_o = r_{1o} + r_{2o} \Rightarrow \dot{r}_o = \dot{r}_{1o} + \dot{r}_{2o}$$

It should be noted again that another condition that must be simultaneously satisfied for these approximations to hold valid is that the angular velocities of the two massive bodies $m_1$ and $m_2$ are small and their distances form the centre of mass are not arbitrarily large, i.e.

$$|\dot{\theta}_1| < 1 \, rad/s, |\dot{\theta}_2| < 1 \, rad/s, |\mathbf{r}_1| << \infty, |\mathbf{r}_2| << \infty$$

Another solvability condition that we require is that $B > 0$. This should be rendered obvious from the definition of $f(t)$ in equation (17). Note also, that the solutions obtained are not valid for the case of collision or explosion analysis. Having discussed in detail, our formal procedure for solution, we now go on to summarize our accomplishments.

## SUMMARY

In this paper we have developed a versatile approach to find approximate solutions for the restricted circular three body problem. The problem finds a lot of applications in celestial mechanics. As mentioned previously, a typical application of the problem would be to describe the motion of an interplanetary probe under the gravitational influence of two massive gravitating bodies, that is planets. We considered the fact that the motion of the two massive bodies $m_1$ and $m_2$ was not effected by the presence or motion of a third body $m_3$ having negligible mass on a relative scale. Using this proposition, we found analytic approximations for the motions of $m_1$ and $m_2$, which were given by the expressions defining $r_1(t)$, $r_2(t)$, $\theta_1(t)$ and $\theta_2(t)$. Of course this was accomplished by a simple consideration of the two body motion executed by $m_1$ and $m_2$. Having done this, we modeled the system of three bodies, taking into account the motion of $m_3$, under the gravitational influence of $m_1$ and $m_2$. The next step was to find approximations for the motion of $m_3$. This was accomplished by using our approximations for the motions of $m_1$ and $m_2$ (found through consideration of two body motion), and performing algebraic manipulations on the system of equations developed while considering three body dynamics.